\documentclass[conference]{IEEEtran}
\usepackage{graphicx}
\usepackage{amsmath}

\long\def\comment#1{}

\def\comment#1{}
\def\0{{\mathbf 0}}

\def\FigSize1{3.5in}

\def\FigSizeFour{3.5in}
\def\0{{\mathbf 0}}

\def\comment#1{}

\begin{document}

\title{Stopping Criteria for Iterative Decoding based on Mutual Information}

%\author{Jinhong~Wu,~\IEEEmembership{Student~Member,~IEEE,}
%        and~Branimir~R.~Vojcic,~\IEEEmembership{Senior~Member,~IEEE}
%
%%\thanks{This work was presented in
%%part at the 2009 IEEE Conference on Information Science and Systems
%%(CISS2009).}
%
%\thanks{J. Wu and B. Vojcic are with the  Department of Electrical and Computer
%Engineering, George Washington University, Washington, DC, USA
%(e-mail: johnnywu@gwu.edu, vojcic@gwu.edu).}}

%\author{Jinhong~Wu,~\IEEEmembership{Member,~IEEE,}
%        and~Branimir~R.~Vojcic,~\IEEEmembership{Senior~Member,~IEEE,} and~Jia~Sheng,~\IEEEmembership{Student~Member,~IEEE,}
%
%\thanks{J. Wu was with Department of Electrical and Computer
%Engineering, George Washington University, Washington, DC, USA. He
%is now with Samsung Information Systems America, San Diego, CA, USA.
%B. Vojcic and J. Sheng are with Department of Electrical and
%Computer Engineering, George Washington University, Washington, DC,
%USA (e-mail: jinhongwu@ieee.org, vojcic@gwu.edu,
%jiasheng@gwu.edu).}}

\author{\authorblockN{Jinhong Wu\\}
\authorblockA{Mobile Solutions lab\\
Samsung Information Systems America\\
San Diego, CA 92121\\
Email: jinhongwu@ieee.org} \and
\authorblockN{Branimir R. Vojcic and Jia Sheng\\}
\authorblockA{ECE Department\\
The George Washington University\\
Washington, DC 20052\\
Email: vojcic@gwu.edu, jiasheng@gwu.edu}}

\maketitle

\begin{abstract}
In this paper we investigate stopping criteria for iterative decoding from a mutual information perspective. We introduce new iteration stopping rules based on an approximation of
the mutual information between encoded bits and decoder
soft output. The first type stopping rule sets a threshold value directly on the approximated mutual information for terminating decoding. The threshold can be adjusted according to the expected bit error rate. The second one adopts a strategy similar to that of the well known cross-entropy stopping rule by applying a fixed threshold on the ratio of a simple metric obtained after each iteration over that of the first iteration. \comment{The convergence properties of
this mutual information are examined and }Compared with several well
known stopping rules, the new methods achieve higher efficiency.

\end{abstract}


\section{Introduction}\label{sec.intro}

Capacity approaching error correction coding schemes such as Turbo codes and LDPC codes are widely adopted in wireless standards, e.g.,
Turbo codes in $3$GPP High Speed Packet Access (HSPA) and Long-Term Evolution (LTE) \cite{3G2585} \cite{3G36212}, LDPC codes in WiMax \cite{IEEE802_16e} and Wi-Fi \cite{IEEE802_11n}. Iterative decoding is the practical solution for decoder implementation in modem chipsets. In order to achieve longer battery life and higher throughput, it is necessary to minimize chipset power consumption and processing delay. Iteration stopping rules serve such purposes by reducing the number of decoding iterations while maintaining the performance. In practice, cyclic redundancy check (CRC) is often employed for error detection which provides an easy solution for early stopping. However, not all systems have CRC at physical layer. For example, HSPA provides a $24$-bit CRC for each transport block but no CRC for each code block within a transport block. A stopping criterion avoids decoding with the maximum number of iterations for every code block. Furthermore, while long CRC (e.g., $32$-bit CRC) requires higher overhead, short CRC (e.g., $8$-bit) results in weaker error detection. Iteration stopping without CRC is therefore of practical interest. In the following, we consider Turbo decoding, but the methodology applies as well to LDPC decoding \cite{Li_You_JingLi_06_ieeecl}.

Iteration stopping has been studied since early days of Turbo codes. The well known cross-entropy (CE)
stopping rule \cite{haop96} uses the relative information between
the two constituent decoders' soft output as the criterion. Decoding
is considered as converged and stopped when this
relative information is close to zero. Based on the same concept,
\comment{the sign change ratio (SCR) rule} two simplified variants
of the CE rule were introduced in \cite{shlf99}: The first one, sign change ratio (SCR) rule, counts
the number of sign changes in the extrinsic log likelihood ratios
(LLRs) between two consecutive iterations. Decoding is terminated when the ratio is small enough. The second one, hard-decision-aided (HDA) rule, compares second decoder output hard decisions with those by the pervious iteration. Decoding stops if all hard decisions remain the same. The overall
performance by the
simplified variants are close to those obtained by the original CE
rule. Further variants or improvements include: sign difference ratio (SDR) \cite{Wu_Woerner_Ebel00} extends from SCR by comparing sign changes of each component decoder's a priori LLRs and extrinsic LLRs; improved hard-decision-aided (IHDA) rule \cite{Ngatched_Takawira_01_eleclett} modifies HDA to compare hard decisions of two component decoders. Aside from savings in memory, the latter two variants perform soft/hard decision comparisons after every half iteration while previous methods perform comparisons after each complete iteration. While other stopping rules are also found in the literature,
in e.g., \cite{Boutillon_Douillard_Montorsi_07_ieeeProc} and
references therein, it is interesting to notice that, above CE, SCR, SDR, HDA, and IHDA stopping rules all originate from the cross-entropy or relative information perspective of decoding convergence.

\comment{The sign change ratio (SCR) rule counts
the number of sign changes in the extrinsic log likelihood ratios
(LLRs) between two consecutive iterations, which is a simplified way
to monitor the CE value.}

%On the other hand, mutual information based convergence analysis has
%been studied in the form of the extrinsic information transfer
%(EXIT) analysis.

In this paper, we consider decoding convergence estimation from a mutual
information perspective. In fact, mutual information analysis has been exemplified by the popular\comment{ Specifically, we examine the
formulation of the popular extrinsic information transfer (EXIT)
analysis} extrinsic information transfer (EXIT) chart \cite{Brink_EXIT_01}. In EXIT analysis, the mutual information between LLRs and the transmitted bit is calculated for system performance evaluation. While it is designed as an off-line tool, a similar and simplified process may be considered for online convergence estimation. In this paper, we show that instead of using true transmitted bits, a simple
approximation of the mutual information by using decoder's hard
decision can be used for efficient iteration stop.

The paper is organized as follows. The system model is described in Section \ref{sec.system_model} along with the approximated calculation of mutual information between encoded bit and LLR. The first iteration stopping rule is formulated in Section \ref{sec.MIA_I}. The second stopping rule is presented in Section \ref{sec.MIA_II}. Simulations are shown in sub-sections \ref{sec:simu-I} and \ref{sec.simu}. Finally, we conclude in Section \ref{sec.conclusions}.
\comment{In fact, analysis on mutual information evolution from iterative decoding has been applied intensively in the literature for performance evaluation. The popular extrinsic information transfer (EXIT) chart is  example.}

\section{System Model and An approximate mutual information calculation}\label{sec.system_model}

%Mutual information between information bit and the log likelihood ratio (LLR), $\Lambda$, generated by a soft-input soft-output (SISO) decoder is defined as
%
%\begin{equation}\label{eq:mutualinfo_def}
%\begin{split}
%&I_{A}=I(u,\Lambda)\\
%&\hspace{0.18in}=1-E_{\Lambda|\mu_{\lambda}}\{\log_{2}[1+\exp(-\Lambda_{A})]\}=\\
%&\frac{1}{\sqrt{2\pi}\sigma_{\lambda}}\int_{-\infty}^{\infty}\exp(-\frac{(\lambda-\mu_{\lambda})^2}{2\sigma_{\lambda}^{2}})
%(1-\log_{2}[1+\exp(-\lambda)])d\lambda.
%\end{split}
%\end{equation}
We consider parallel concatenated convolutional code (PCCC) \cite{begt93} where coded bits are mapped to symbols from a signal constellation and transmitted over a memoryless channel with additive white Gaussian noise (AWGN). At the receiver, assume perfect channel knowledge and optimal demodulation, iterative Log-MAP decoding \cite{bcjr74} \cite{haop96}is performed. \comment{By above optimality assumption we obtain channel LLRs, a priori LLRs, as well as extrinsic LLRs for iterative decoding that satisfy the consistency condition \cite{Hagenauer_04_EuSIPCO} \cite{TCOM01_ten_Brink}. By satisfying the consistency condition, the LLRs provide true probabilities of each bit's value being $1$ or $0$\footnote{In real systems, imperfect channel information and sub-optimal processing may generate LLRs that do not strictly satisfy the consistency condition. However, an approximation of such assumption is still practical.}.} The mutual information between each information bit $u\in\{1,-1\}$ and its associated LLR $\Lambda$ can be derived as \cite{Brink_EXIT_01}

\begin{equation}\label{eq:mutualinfo_Guassian}
%\begin{split}
I=1-E_{\Lambda|\mu_{\lambda}}\{\log_{2}[1+\exp(-\Lambda)]\}.
%&\frac{1}{\sqrt{2\pi}\sigma_{\lambda}}\int_{-\infty}^{\infty}\exp(-\frac{(\lambda-\mu_{\lambda})^2}{2\sigma_{\lambda}^{2}})
%(1-\log_{2}[1+\exp(-\lambda)])d\lambda.
%\end{split}
\end{equation}
where $\Lambda$ denotes LLR in general, from which we may further specify a priori LLR, $\Lambda_{a}$, extrinsic LLR, $\Lambda_{e}$, and a posteriori LLR $\Lambda_{app}$ by adding appropriate subscripts.
%\begin{equation}\label{eq:mutualinfo_Guassian}
%\begin{split}
%&I_{A}=I(u,\Lambda_{A})=\\
%&\frac{1}{\sqrt{2\pi}\sigma_{\lambda}}\int_{-\infty}^{\infty}\exp(-\frac{(\lambda-\mu_{\lambda})^2}{2\sigma_{\lambda}^{2}})
%(1-\log_{2}[1+\exp(-\lambda)])d\lambda\\
%&\hspace{0.18in}=E_{\Lambda|\mu_{\lambda}u}\{1-\log_{2}[1+\exp(-\Lambda_{A})]\}.
%\end{split}
%\end{equation}

By applying the ergodicity assumption on LLR distribution,
\eqref{eq:mutualinfo_Guassian} is simplified as
\cite{Hagenauer_04_EuSIPCO}
\begin{equation}\label{eq:I_appr}
I\approx 1-\frac{1}{N}\sum\limits_{n=1}^{N}\log_{2}[1+\exp(-u_{n}\Lambda(n))]\\
\end{equation}
where $\Lambda(n)$ refers to the a priori LLR for bit $u_{n}$,
$n=1,2,\cdots,N$.

%Similarly, mutual information between the
%information bit and decoder output extrinsic LLR, $\Lambda_{e}$, can
%be approximated by
%
%\begin{equation}\label{eq:IE_appr}
%I_{e}\approx 1-\frac{1}{N}\sum\limits_{n=1}^{N}\log_{2}[1+\exp(-u_{n}\Lambda_{e,n})]\\
%\end{equation}

%When the output extrinsic LLR, $\Lambda_{E}$, may not be Guassian
%distributed, an empirical probability distribution function,
%$P_{E}(\Lambda_{E}|u)$, can be measured to evaluate the corresponding mutual information. The output mutual information can be calculated as \cite{Brink_EXIT_01}
%
%\begin{equation}\label{eq:mutualinfo_Other}
%\begin{split}
%&I_{E}=\\
%&\frac{1}{2}\sum\limits_{u=\pm1}\int_{-\infty}^{\infty}P_{E}(\xi|u)\log_{2}(\frac{2P_{E}(\xi|u)}{P_{E}(\xi|U=-1)+P_{E}(\xi|U=1)})d\xi.
%\end{split}
%\end{equation}

Convergence analysis based on above mutual
information evolution has been utilized by the well-known EXIT chart \cite{Brink_EXIT_01}
\cite{Hagenauer_04_EuSIPCO}. It takes the mutual information between
the information bit and the associated a priori LLR, $I_{a}$, as the input, and generates the mutual information
between the information bit and its extrinsic LLR, $I_{e}$, as the
output. \comment{A high output mutual information indicates a high
reliability from the soft information, and vice versa.} An EXIT chart is generated by plotting the output $I_{e}$ values
corresponding to a sequence of input $I_{a}\in[0,1]$. A high mutual
information indicates high reliabilities of LLRs, and vice versa.

Note that \eqref{eq:I_appr} requires knowledge of the information
bits and therefore is used for off-line analysis. A blind mutual information calculation is presented in \cite{Land_Hoeher_Gligorevic_04_ITG_SourceChannelCoding}. However, it requires more computations. For simple online analysis, we use the decoder hard decisions as
estimates of the information bits, which results in an approximation
of the desired mutual information $I$.
Clearly, its accuracy depends on the reliability of the hard
decisions. However, since output LLRs provides probabilistic measure
about information bits, this approximation can approach the maximum
value only if LLRs are large enough, which in turn indicates the
decoding convergence
\cite{Reid_Gulliver_Taylor_01_ConvergenceError}. Therefore, for the
purpose of iteration stopping, a high threshold value on the
approximated mutual information can be effective. On the other hand,
we are interested in knowing the reliability of the overall decoder
output, rather than the extrinsic information alone. Therefore we
may consider the mutual information generated by
a posteriori LLR, $\Lambda_{app}$. In fact, using
$\Lambda_{app}$ instead of $\Lambda_{e}$ allows for earlier
identification of decoding convergence.

Applying the hard decision by $\hat{u}=sign(\Lambda_{app})$, we approximate
\eqref{eq:I_appr} of a posteriori LLR, $I_{app}$, by

\begin{equation}\label{eq:MI_online_appr_II}
\begin{split}
%\hat{I}_{app}&=I(\hat{u},\Lambda_{app})\\
%&\approx 1-E_{\Lambda_{app}|\mu_{\lambda_{app}},u=\hat{u}}\{\log_{2}(1+\exp(-\Lambda_{app}))\}\\
&\hat{I}_{app}\\
&\approx\frac{1}{N}\sum\limits_{n=1}^{N}(1-\log_{2}(1+\exp(-\Lambda_{app}(n)\cdot sign(\Lambda_{app}(n))))) \\
&=1-\frac{1}{N\ln(2)}\sum\limits_{n=1}^{N}\ln(1+e^{-|\Lambda_{app}(n)|})\\
&\approx1-\frac{1.44}{N}\sum\limits_{n=1}^{N}\ln(1+e^{-|\Lambda_{app}(n)|})\\
%&=1-\frac{1}{N\ln(2)}\sum\limits_{n=1}^{N}\Epsilon\\
&=1-\epsilon
\end{split}
\end{equation}
where we define \comment{$\Epsilon=\ln(1+e^{-|\Lambda_{app}(n)|})$
and }
$\epsilon=\frac{1.44}{N}\sum\limits_{n=1}^{N}\ln(1+e^{-|\Lambda_{app}(n)|})$.

Furthermore, adopting the common practice in the log-MAP algorihtm
implementations \cite{haop96}, the calculation of
$\ln(1+\exp(-|A|))$ can easily handled by a look up table function,
$LUT(|A|)$.\comment{using the Jacobian identity \cite{rovh95}
\begin{equation}\label{eq_Jacobian}
\ln(\exp(a)+\exp(b))=max(a,b)+\ln(1+\exp(-|a-b|)),
\end{equation}} Therefore, \eqref{eq:MI_online_appr_II} is simplified as

\begin{equation}\label{eq:MI_online_appr_II}
\hat{I}_{app}\approx1-\frac{1.44}{N}\sum\limits_{n=1}^{N}LUT(|\Lambda_{app}(n)|).
\end{equation}

\begin{figure}
\centering
\includegraphics[width=\FigSizeFour]{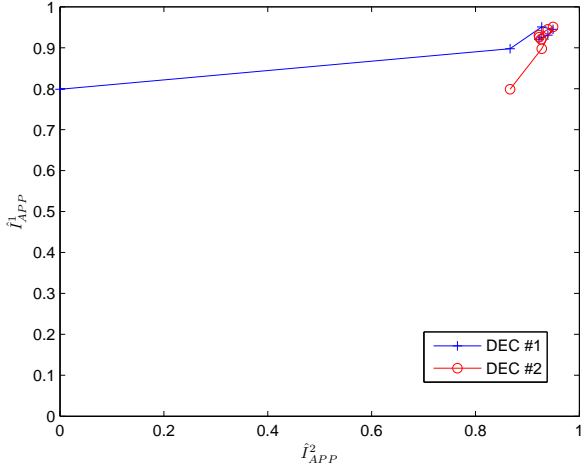}
\caption{Evolution of the approximated mutual information: Decoding convergence not reached ($E_b/N_0=1dB$).}
\label{fig_MI_nonConverge}
\end{figure}

\begin{figure}
\centering
\includegraphics[width=\FigSizeFour]{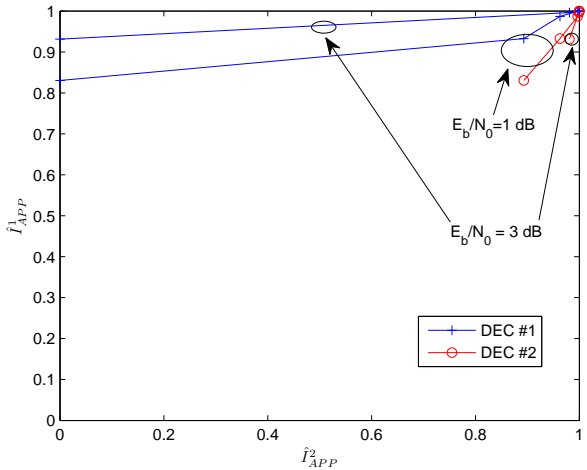}
\caption{Evolution of the approximated mutual information: Decoding convergence reached at low SNR ($E_b/N_0=1dB$) and high SNR ($E_b/N_0=3dB$).}
\label{fig_MI_Converge_lowSNR_highSNR}
\end{figure}

Note that due to the averaging over all bit decisions, this approximation becomes more accurate as the iterative
decoding converges to the reliable decision. When the decoding delivers reliable results, $\Lambda_{app}(n)$
values are generally high. Consequently, $\hat{I}_{app}\rightarrow1$
by \eqref{eq:MI_online_appr_II}. In contrast, erroneous decoding
will result in a lower $\hat{I}_{app}$ value due to smaller magnitudes of LLRs. For illustration purpose, three typical examples
are shown in Fig.\ref{fig_MI_nonConverge} -
\ref{fig_MI_Converge_lowSNR_highSNR}: They are
based on decoding a ($7$, $5$) code using random interleavers of
length $900$. Coded data are binary phase shift keying (BPSK) modulated and transmitted over AWGN channel. Fig.~\ref{fig_MI_nonConverge} is obtained from
unsuccessful decoding of a packet at $E_{b}/N_{0}=1dB$.
The inner two curves in Fig.~\ref{fig_MI_Converge_lowSNR_highSNR} are from successful decoding of
another packet also at $E_{b}/N_{0}=1dB$, while the outer two curves in
Fig.~\ref{fig_MI_Converge_lowSNR_highSNR} are from successful decoding of a
packet at $E_{b}/N_{0}=3dB$. The evolution of $\hat{I}_{app}$ during
the iterations is plotted: The vertical axis, denoted by
$\hat{I}^{1}_{app}$, refers to the mutual information produced by
the $1^{st}$ constituent decoder. The horizontal axis, denoted by
$\hat{I}^{2}_{app}$, refers to that produced by the $2^{nd}$
constituent decoder. By sequential exchange of extrinsic
information, each decoder accepts the mutual information produced by
the other decoder as the input, and generates its own mutual
information as the output. The two curves in each plot indicate how
each decoder's output mutual information evolves according to the
input mutual information it receives from the other decoder.

%\begin{figure}
%\centering
%\includegraphics[width=\FigSizeFour]{MI_Converge_lowSNR}
%\caption{Evolution of mutual information between LLR and hard
%decision: Decoding convergence reached at low SNR ($E_b/N_0=1dB$).}
%\label{fig_MI_Converge_lowSNR}
%\end{figure}
%%Rate $1/2$, $(7,5)$ turbo code, interleaver length $900$
%
%\begin{figure}
%\centering
%\includegraphics[width=\FigSizeFour]{MI_Converge_highSNR}
%\caption{Evolution of mutual information between LLR and hard
%decision: Decoding convergence reached at high SNR ($E_b/N_0=3dB$).}
%\label{fig_MI_Converge_highSNR}
%\end{figure}

\section{$\epsilon$ threshold and type-I Stopping rule}\label{sec.MIA_I}

Based on the characteristics shown above, we propose a mutual
information aided (MIA) iteration stopping rule: decoding stops when
$\hat{I}^{2}_{app}$ is very close to $1$, i.e., $\epsilon$ is small
enough. \footnote{It is reasonable to apply the threshold on
$\hat{I}_{app}$ also for $1$st decoder, which will result in
decoding stops $0.5$ iterations earlier for some cases.}
\comment{The computational overhead for this mutual information is
lower than that required by the CE rule. } We denote this stopping rule as type-I mutual information aided (MIA-I) rule. For simplicity, a fixed
threshold value on $\epsilon$ (e.g., $10^{-5}$) may be applied.\comment{ Such
an approach has been investigated in
\cite{Wu_Vojcic_Sheng_2011Asilomar}.} For flexible iteration stopping, more specifics about $\epsilon$ are considered as follows.

\subsection{Estimation of Bit error rate and $\epsilon$ threshold}

According to the definition of LLR, it is straightforward to derive the conditional bit error probability given the a posteriori LLR $\Lambda$ as

\begin{equation}\label{BER_estimate}
    P_{e|\Lambda} = \frac{1}{1+e^{|\Lambda|}}.
\end{equation}
For large $|\Lambda|$ it then follows that
\begin{equation}\label{BER_epsilon_approx}
    P_{e|\Lambda}=\frac{1}{1+e^{\Lambda}}\approx \frac{1}{e^{\Lambda}}\approx\ln(1+e^{-|\Lambda|}).
\end{equation}
For reasonably large $|\Lambda|$ (e.g., for $|\Lambda|\geq 2$), we can verify that the difference from approximation is negligible.
%$\ln(1+e^{-|\Lambda|})\approx \frac{1}{e^{\Lambda}}\approx\frac{1}{1+e^{\Lambda}}=P_{e}$ with negligible difference.
Therefore $\epsilon$ in \eqref{eq:MI_online_appr_II} could also be
used for a rough estimation of decoding bit error rate (BER). Note that LLRs
produced by early iterations are more accurate but less so in later
iterations due to increased correlations with soft input to the
decoder(s) \cite{Brink_EXIT_01}\comment{\cite{ElGamal_Hammons_GuassianAppr_01}}. For this reason,
$\epsilon$ approximates BER better in early iterations but not as
well in later iterations (as it usually approaches zero with a large
number of iterations). \comment{Consequently, it is valid to
determine a proper $\epsilon$ value according to desired BER. As a
result, a valid $\epsilon$ threshold should not exceed achievable
BER significantly. In other words,}For effective iteration stopping,
the threshold $\epsilon$ may be selected to be around the value of
expected BER. Consider a fixed threshold of $\epsilon$, it is intuitively clear that a lower threshold ensures lower BER but may require more iterations. This provides the following two options of selecting a threshold of  $\epsilon$.

Option A: When decoding throughput is of higher priority but BER is less important, a relatively high threshold of $\epsilon$ can be applied. This may be appropriate for certain real time applications such as voice or video transmissions with certain quality of service (QoS) requirement. In some practical scenarios BER of $10^{-3}$ to $10^{-4}$ can be acceptable. In those scenarios, a fixed threshold on $\epsilon$ may be chosen from, e.g., $10^{-2}$ to $10^{-5}$, for the desired BER. For notation purpose we denote MIA-I rule with fixed $\epsilon$ threshold by MIA-I-A.

Option B: When decoding reliability is of higher priority, an estimation of achievable BER can be used for $\epsilon$ threshold. For known channel and coding scheme, BER can be measured in advance. If such a BER value is unknown a priori, a sufficiently low $\epsilon$ may be used initially, and then adjusted
accordingly once BER can be measured or estimated. This may apply to channels with slow changes. We denote MIA-I rule with such adaptive $\epsilon$ thresholds by MIA-I-B.

\subsection{Simulations}\label{sec:simu-I}
The effects of different thresholds are illustrated by simulations. The PCCC encoding scheme uses two identical ($7$,$5$) component encoders, with random interleaving of size $900$. Puncturing of even (odd) indexed parity bits by first (second) encoder is applied to generate the rate $1/2$ code. Coded data are modulated with BPSK and transmitted over AWGN channel.

For MIA-I-A, we applied $\epsilon=10^{-2}$, $\epsilon=10^{-3}$, and $\epsilon=10^{-4}$ for different trade-offs between BER and number of iterations. For MIA-I-B, we adaptively set $\epsilon$ according to Table~\ref{tb_thershold_MIA_I_A} to match the achievable BER. We compare these MIA stopping rules with the CE rule as well as the HDA rule \footnote{Other stopping methods, e.g., SCR, SDR or IHDA, etc, are similar to or slightly worse than CE or HDA in performance and/or average number of iterations as reported in \cite{shlf99}, \cite{Wu_Woerner_Ebel00} and \cite{Ngatched_Takawira_01_eleclett}.} \footnote{Also note that the proposed MIA rules can also compare $\hat{I}_{app}$ or $\epsilon$ ratios after each half iteration with trivial modification, which could further reduce the average iteration number.}. The threshold for the CE criterion is $10^{-4}$. All stopping rules are then compared with decoding using $6$ iterations.

\begin{figure}
\centering
\includegraphics[width=\FigSizeFour]{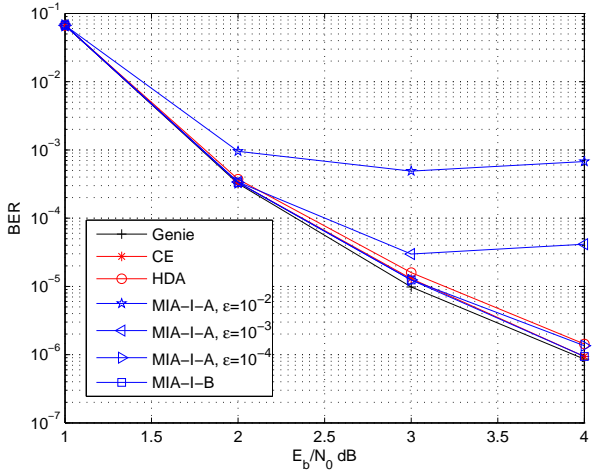}
\caption{Performance by different $\epsilon$ threshold: rate $1/2$ turbo code with memory length $2$,
interleaver size $900$, over AWGN channel.}
\label{fig_BER900_epsilonA234_B}
\end{figure}

\begin{figure}
\centering
\includegraphics[width=\FigSizeFour]{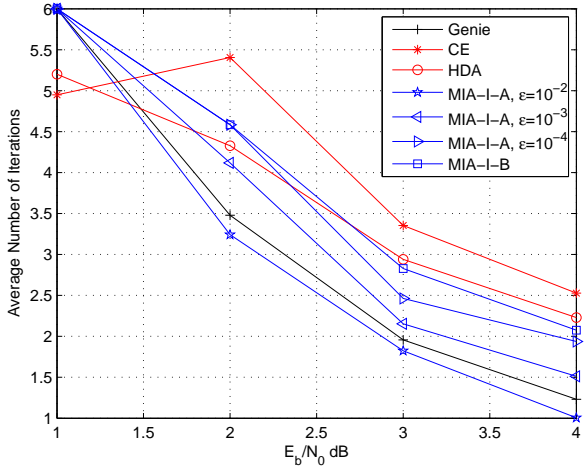}
\caption{Average number of iterations by different $\epsilon$ threshold: rate $1/2$ turbo code with memory length $2$,
interleaver size $900$, over AWGN channel.}
\label{NumIter900_epsilonA234_B}
\end{figure}

We plot BER curves as well as the average number of iterations in Fig.~\ref{fig_BER900_epsilonA234_B} and Fig.~\ref{NumIter900_epsilonA234_B}.
As a reference, BER as well as the average numbers of iterations assuming a `genie'
error detector are also plotted. Its number of iterations
refers to either the lowest number of iterations for error free
decoding, or the maximum number of $6$ iterations if errors always
exist.

\begin{table}
\renewcommand{\arraystretch}{1.3}
\caption{Threshold of $\epsilon$ for different SNR} \label{tb_thershold_MIA_I_A}
\begin{center}
\begin{tabular}{c}
(7,5) code with interleaver size 900, AWGN channel
\end{tabular}
\end{center}
\begin{center}
\begin{tabular}{|c|c|c|c|c|}
  \hline
  $E_b/N_0$ & 1 & 2 & 3 & 4 \\
  \hline
$\epsilon$ & $10^{-2}$ & $10^{-4}$ & $10^{-5}$ & $10^{-6}$ \\
  \hline
\end{tabular}
\end{center}
\end{table}

\section{$\epsilon$ ratio threshold and type-II Stopping rule}\label{sec.MIA_II}

To avoid estimation of BER for determination of $\hat{I}_{app}$ threshold, we further consider the metric $\epsilon$. As shown by Fig.~\ref{fig_MI_Converge_lowSNR_highSNR}, decoding convergence accompanies the minimization of $\epsilon$. Interestingly, in average $\epsilon(iter)$ shows a similar pattern as that of the CE criterion \cite{haop96} that we can leverage.

To illustrate we briefly review the CE rule below. Denote iteration number as $iter$ and consider extrinsic and a posteriori LLRs output from second component decoder , CE is expressed as \cite{haop96}
\begin{equation}\label{eq:CE_defination}
  CE(iter)=\frac{1}{N}\sum\limits_{n=1}^{N}\frac{|\triangle\Lambda^{iter}_{e}(n)|^2}{e^{\Lambda^{iter}_{app}(n)}}
\end{equation}
where $\triangle\Lambda^{iter}_{e}=\Lambda^{iter}_{e}-\Lambda^{iter-1}_{e}$.

The effectiveness of CE rule comes from a fact that the ratio of $\frac{CE(iter)}{CE(1)}$ typically shows an accelerated decreasing as decoding converges. Decoding stops if $\frac{CE(iter)}{CE(1)}<10^{-3}$ or $\frac{CE(iter)}{CE(1)}<10^{-4}$. Usually choosing $10^{-3}$ saves slightly in iteration numbers but more often result in an early error floor, while $10^{-4}$ usually maintain performance better at the cost of slightly higher number of iterations.

\begin{figure}
\centering
\includegraphics[width=\FigSizeFour]{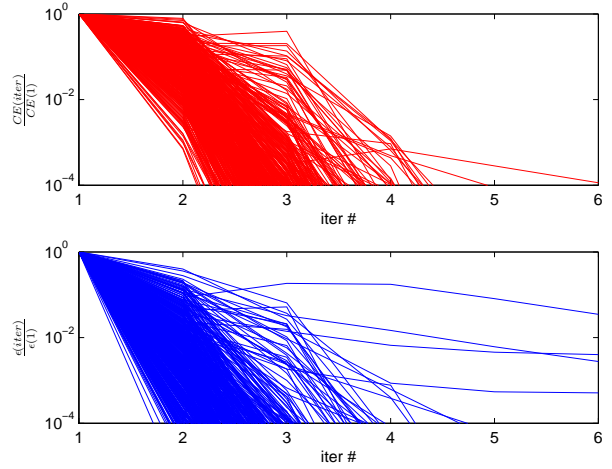}
\caption{cross-entropy ratio and $\epsilon$ ratio over iterations: rate $1/2$ turbo code with memory length $2$,
interleaver size $900$, over AWGN channel.}
\label{fig_CEration_MIratio_3dB_A}
\end{figure}

To show the pattern of $\frac{\epsilon(iter)}{\epsilon(1)}$ in comparison to $\frac{CE(iter)}{CE(1)}$, we plot numerically such two ratios over iterations in Fig.~\ref{fig_CEration_MIratio_3dB_A}. The figures are randomly generated by decoding $500$ coded packets at $E_{b}/N_{0}=3$ dB and measuring those two ratios by each packet after each iteration. In both figures one curve is plotted for each decoded packet. In average, $\frac{\epsilon(iter)}{\epsilon(1)}$ drops faster. This implies the possibility of earlier stopping compared with the CE rule.

Based on the observation, we express the second stopping criterion as $\frac{\epsilon(iter)}{\epsilon(1)}<10^{-3}$. We denote this stopping rule as type-II mutual information aided (MIA-II) rule. We further note that computation-wise, $\epsilon(iter)$ requires $N$ look-up table search, $N-1$ additions, and one multiplication, while $CE(iter)$ requires $N$ times look-up table search (for the exponential function), $2N-1$ additions, and $2N+1$ multiplications.

\subsection{Simulations}\label{sec.simu}

In this section we compare MIA-II and MIA-I-B with the CE rule as well as the HDA rule\comment{ \footnote{Other stopping methods, e.g., SCR, SDR or IHDA, etc, are similar to or slightly worse than those two in performance and/or average number of iterations as reported in \cite{shlf99}, \cite{Wu_Woerner_Ebel00} and \cite{Ngatched_Takawira_01_eleclett}.} \footnote{Also note that the proposed MIA rules can also compare $\hat{I}_{app}$ or $\epsilon$ ratios after each half iteration with trivial modification, which could further reduce the average iteration number.}}.
Simulations are based on PCCC, with the ($7$, $5$) component
code with random interleavers of size $2048$. We compare different stopping rules for transmissions by BPSK over AWGN channel and fast Rayleigh fading channel.

\begin{figure}
\centering
\includegraphics[width=\FigSizeFour]{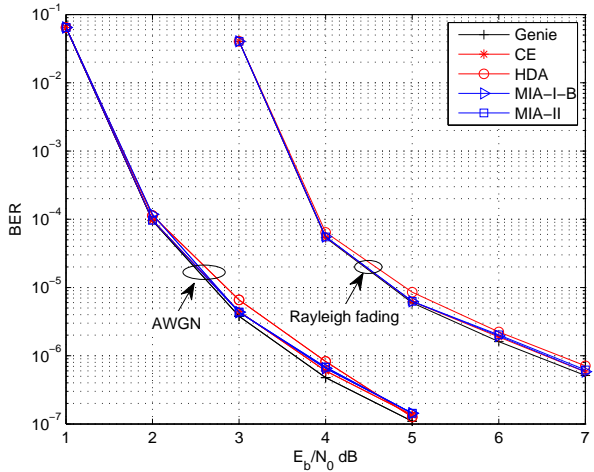}
\caption{BER: rate $1/2$ turbo code with memory length $2$,
interleaver size $2048$, over AWGN channel and fast Rayleigh fading channel.}
\label{fig_BER2048_A_F_6iter}
\end{figure}

\begin{figure}
\centering
\includegraphics[width=\FigSizeFour]{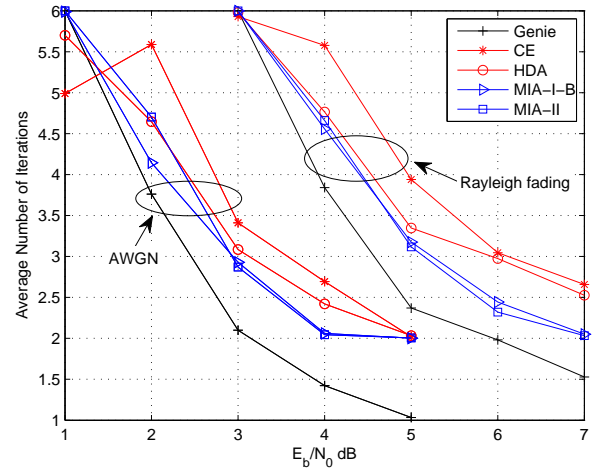}
\caption{Average number of iterations: rate $1/2$ turbo code with
memory length $2$, interleaver size $2048$, over AWGN channel and fast Rayleigh fading channel.} \label{fig_NumIter2048_A_F}
\end{figure}

For MIA-I-B stopping rule, the threshold value is chosen as in Table~\ref{tb_thershold}. \comment{For comparison, a fixed threshold
of $\epsilon=10^{-5}$ is also tested. }
The BER performance comparison is shown in Fig.~\ref{fig_BER2048_A_F_6iter}\comment{ and Fig.~\ref{fig_BER2048_F_6iter}}. The average numbers of
iterations by each rule are shown in Fig.~\ref{fig_NumIter2048_A_F}\comment{ and Fig.~\ref{fig_NumIter2048_F}}. \comment{The simulations show that although the MIA stopping
rule incurs a slightly higher error floor, the overall efficiency is
evidently higher than the CE rule. The average number of iterations
by MIA is about $1$ to $1.5$ less than that by the CE rule for most
SNRs. } Compared with the CE rule or the HDA rule, the proposed MIA rules usually
stops earlier.

%On the other hand, we also see that, when applying
%the fixed $\epsilon$ threshold, the decoder stops after more
%unnecessary iterations at low SNRs, while resulting an early error
%floor at high SNRs.

%For more comparisons, we test with memory $3$ ($15$, $13$) code, and
%memory $4$ ($23$, $35$) code. The comparisons are shown in Fig.
%\ref{fig_BER900_mem34code_6iter}, \ref{fig_NumIter900_15_13} and
%Fig. \ref{fig_NumIter900_23_35}.

\begin{table}
\renewcommand{\arraystretch}{1.3}
\caption{$\epsilon$ Threshold values for different SNR} \label{tb_thershold}
%\begin{center}
%\begin{tabular}{c}
%(7,5) code with interleaver size 900
%\end{tabular}
%\end{center}
%\begin{center}
%\begin{tabular}{|c|c|c|c|c|c|}
%  \hline
%  $E_b/N_0$ & 1 & 2 & 3 & 4 & 5 \\
%  \hline
%$\epsilon$ & $10^{-2}$ & $10^{-4}$ & $10^{-5}$ & $10^{-6}$ & $10^{-7}$ \\
%  \hline
%\end{tabular}
%\end{center}
\begin{center}
\begin{tabular}{c}
(7,5) code with interleaver size 2048, AWGN channel
\end{tabular}
\end{center}
\begin{center}
\begin{tabular}{|c|c|c|c|c|c|}
  \hline
  $E_b/N_0$ & 1 & 2 & 3 & 4 & 5 \\
  \hline
$\epsilon$ & $10^{-1}$ & $10^{-3}$ & $10^{-5}$ & $10^{-6}$ & $10^{-7}$ \\
  \hline
\end{tabular}
\end{center}

\begin{center}
\begin{tabular}{c}
(7,5) code with interleaver size 2048, Rayleigh fading channel
\end{tabular}
\end{center}
\begin{center}
\begin{tabular}{|c|c|c|c|c|c|}
  \hline
  $E_b/N_0$ & 3 & 4 & 5 & 6 & 7 \\
  \hline
$\epsilon$ & $2\times10^{-2}$ & $2\times10^{-4}$ & $2\times10^{-5}$ & $5\times10^{-6}$ & $10^{-6}$ \\
  \hline
\end{tabular}
\end{center}

%\begin{center}
%\begin{tabular}{c}
%(15,13) code with interleaver size 900
%\end{tabular}
%\end{center}
%\begin{center}
%\begin{tabular}{|c|c|c|c|c|}
%  \hline
%  $E_b/N_0$ & 1 & 2 & 3 & 4 \\
%  \hline
%$\epsilon$ & $10^{-2}$ & $10^{-5}$ & $10^{-6}$ & $10^{-7}$ \\
%  \hline
%\end{tabular}
%\end{center}
%\begin{center}
%\begin{tabular}{c}
%(23,35) code with interleaver size 10000
%\end{tabular}
%\end{center}
%\begin{center}
%\begin{tabular}{|c|c|c|c|}
%  \hline
%  % after \\: \hline or \cline{col1-col2} \cline{col3-col4} ...
%  $E_b/N_0$ & 1 & 2 & 3 \\
%  \hline
%$\epsilon$ & $10^{-2}$ & $10^{-6}$ & $10^{-7}$ \\
%  \hline
%\end{tabular}
%\end{center}

\end{table}

\section{Conclusions}\label{sec.conclusions}

Using turbo code as example, we provided an approximate calculation of the mutual information
between encoded bit and decoder's a posteriori LLR. The
changing pattern of such metric
effectively indicates decoding convergence after iterations. Two types of iteration stopping rules are proposed.
The first type includes two options: MIA-I-A using a fixed threshold provides flexible trade-off between performance and complexity; MIA-I-B ensures the performance by an estimation of the achievable BER. The second type, MIA-II, adopts a more universal threshold with a strategy similar to that of the CE rule, but requires less computation. Simulations show that compared with CE and HDA stopping rules, both MIA-I-B and MIA-II achieve higher efficiency while maintaining achievable performance by a maximum number of iterations.

\bibliographystyle{IEEEtran}

%\bibliographystyle{IEEE}
%\addcontentsline{toc}{chapter}{\numberline{}Bibliography}
\bibliography{IEEEabrv,Jinhong}

\end{document}